\documentclass[prl,aps,twocolumn,showpacs,bibnotes,epsf]{revtex4}
\usepackage{graphicx, verbatim}
\usepackage{amsmath}

\usepackage{epstopdf}
\usepackage{CJK}
\usepackage[utf8]{inputenc}
\usepackage[english]{babel}
\usepackage{hyperref}
\hypersetup{colorlinks=true,linkcolor=blue,filecolor=magenta,urlcolor=cyan}

\begin{document}
\begin{CJK*}{GB}{gbsn}
\title{Physical reservoir computing built by spintronic devices for temporal information processing}
\author{Wencong Jiang$^*$}
\author{Lina Chen\footnote{These authors contributed equally to this work}}
\author{Kaiyuan Zhou}
\author{Liyuan Li}
\author{Qingwei Fu}
\author{Youwei Du}
\author{R.H. Liu\footnote{Corresponding author email: rhliu@nju.edu.cn}}

\affiliation{National Laboratory of Solid State Microstructures, School of Physics and Collaborative Innovation \\Center of Advanced Microstructures, Nanjing University, Nanjing 210093, China}

\begin{abstract}
Spintronic nanodevices have ultrafast nonlinear dynamic and recurrence behaviors on a nanosecond scale that promises to enable spintronic reservoir computing (RC) system. Here two physical RC systems based on a single magnetic skyrmion memristor (MSM) and 24 spin-torque nano-oscillators (STNOs) were proposed and modeled to process image classification task and nonlinear dynamic system prediction, respectively. Based on our micromagnetic simulation results on the nonlinear responses of MSM and STNO with current pulses stimulation, the handwritten digits recognition task domesticates that an RC system using one single MSM has the outstanding performance on image classification. In addition, the complex unknown nonlinear dynamic problems can also be well solved by a physical RC system consisted of 24 STNOs confirmed in a second-order nonlinear dynamic system and NARMA10 tasks. The capability of both high accuracy and fast information processing promises to enable one type of brain-like chip based on spintronics for various artificial intelligence tasks.
\end{abstract}

\pacs{85.75.F+, 07.05.Mh, 75.60.Ch}
\maketitle
\end{CJK*}

\section{I. Introduction}
The deep neural network recently achieved significant success and widespread influence in applications of artificial intelligence (AI), e.g., medical diagnosis, intelligent machines, entertainment, the game of Go\cite{deep1,deep2,GO}. Neuromorphic computing - a new non-von Neumann computing paradigm inspired by how the human brain works that could improve the performance and power efficiency for the future of AI exponentially, has attracted considerable attention recently\cite{prl6,JaegerScience2004,prl7}. In neuromorphic computing, hardware implementations of neural networks are becoming a dominant approach to reach the energy efficiency of the human brain.

Reservoir computing (RC), as a paradigm is well-suited to hardware implementations due to the fixed nature of the weights between neurons in the reservoir model itself, has been studied intensively in theory and experiments\cite{prl6,JaegerScience2004,prl7,AppeltantNC2011,DuNC2017,TorrejonNature2017}. RC only requires the reservoir model should have two properties: sufficiently complex dynamics and fading memory effects\cite{prl6,prl7,JaegerScience2004}. Since the connections among nodes of reservoir network can be hidden and random fixed, the weights inside the reservoir does not need to be trained, only the weights of the last linear output layer need to be modulated or trained\cite{JaegerScience2004, prl7}, RC is very suitable or natural to build a physical reservoir based on the actual physical devices with complex dynamics and short-term memory behaviors\cite{AppeltantNC2011}. Therefore, the physical reservoir has great advantages on neuromorphic computing, for unlike software-based reservoir it does not need colossal computing resource to built and fine-tune the dynamic states due to connections among nodes of reservoir network intrinsically encoded in actual physical devices.

In recent years, a variety of physical devices have been proposed to build reservoir neural network, such as phase-change memories\cite{Eryilmaz2014}, memristors\cite{DuNC2017,PreziosoNature2015,Sheridan2017,GuptaNC2016}, spintronic devices \cite{TorrejonNature2017,Vowel-recognition}, optoelectronics\cite{Paquot2012,optprl} etc., offering new opportunities for neuromorphic computing and AI in hardware design. Spintronics is an emerging technology that exploits electronic charge, the intrinsic spin of the electron and its associated magnetic moment in solid-state devices. Very recently, spin torque (ST) memristor and nano-oscillator devices are proposed and found having huge potential applications in multi-bit data storage and logic and development of artificial synaptic devices\cite{TorrejonNature2017,Vowel-recognition,pra1}. Magnetic memristor encodes information based upon spin torque driven magnetic domain wall (such as skrymion\cite{rhliuPRL2015,Sampaio2013,Wanjun,IEEE2016}) motion. Spin-torque nano-oscillator (STNO), owning the complex non-linear dynamics and relaxation characteristics (short-term memory) from magnetization precession with microwave-frequency caused by spin torque effect, can perform high-speed data processing based on its magnetic dynamics and magnetoresistance effect (MR)\cite{PribiagNaturePhysics2007,rhliuPRL2013,Houshang2016}. Since the complex magnetic dynamics and both high-speed store and process information ability, the spintronic device is an attractive candidate to develop efficient neuromorphic computing systems (non-von Neumann computing) for highly data-centric artificial intelligence related applications \cite{Grollier2016,add1,add4}. Here, we propose two spintronic devices: magnetic skyrmion memristor (MSM) and STNO to build physical RC system for neuromorphic computing and demonstrate that they have the outstanding performance on temporal information processing including image classification and nonlinear dynamics system prediction based on our micromagnetic simulation results.

\begin{figure*}[!t]
\centering
\includegraphics[width=0.8\textwidth]{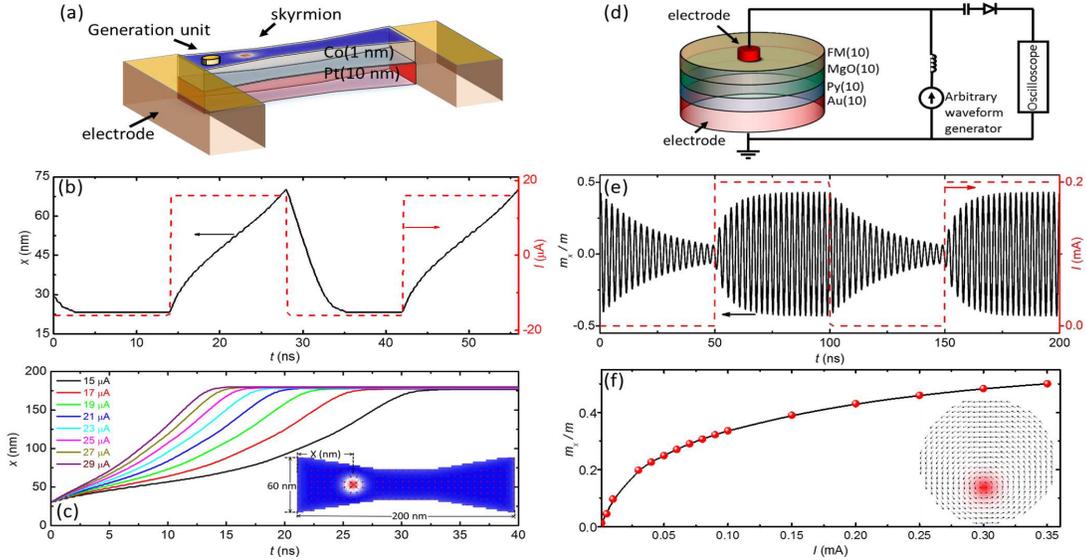}
\caption{Nonlinear dynamics of magnetic skyrmion memristor (MSM) and spin torque nano-oscillator (STNO) under current pulse stimulation. (a) Structure schematic of a MSM consisted of the dumbbell shape ferromagnetic Pt(10)/Co(1) bilayer and Au electrodes. (b) The position $x$ (left axis) of skyrmion under an ac continual square current pulse stimulation (right axis). (c) Time dependent position $x$ at different dc driving currents, as labeled. Inset: the position $x$ of skyrmion was defined in the dumbbell Pt/Co bilayer. (d) Schematic of a vortex-type STNO and blockdiagram. (e) The response of the normalized x-component of magnetization $m_x$ (left axis) under an square current pulse (right axis) with an amplitude of 0.2~mA and pulse width of 50 ns. (f) Dependence of the maximum envelope of $m_x/m$ procession motion at steady state on the input current $I$.} \label{fig1}
\end{figure*}

\section{II. Modelling and simulating physical reservoirs}

\subsection{A. Magnetic skyrmion memristor}

To construct a physical RC system for neuromorphic computing using actual devices, the reservoir should satisfy enough variations and the echo property meaning the current state should be able to feel the influence of its history\cite{JaegerScience2004}. Both complex nonlinear dynamics and short-term memory properties make spintronic nanodevices as an outstanding candidate to construct a physical reservoir. Since micromagnetic simulation currently recognized as an indispensable tool in the field of magnetism research, the nonlinear dynamics and memory properties of spin-torque nanodevices are obtained through solving the Landau-Lifshitz-Gilbert (LLG) equation using the OOMMF code\cite{kronmuller}.
Fig. 1(a) shows the structure schematic of the magnetic skyrmion memristor based on the dumbbell shape ferromagnetic Pt(10 nm)/Co(1 nm) bilayer with perpendicular magnetic anisotropy (PMA). To further enhance the nonlinear behavior of skyrmion motion, the dumbbell shape of 60 $\times$ 200 nm with 30 nm center width is selected, as shown in the inset of Fig. 1(c). The material parameters used in the simulations are the exchange stiffness $A$ = 15 pJ/m, the saturation magnetization $M_{s}$ = 580 kA/m, the perpendicular anisotropy $K_{u}$ = 0.7 MJ/m$^3$, the Dzyaloshinskii-Moriya interaction(DMI) $D$ = 2.9 mJ/m$^3$, the Gilbert damping constant $\alpha$ = 0.3, and the spin Hall angle $P$ = 0.1. The cell size of 2 nm $\times$ 2 nm $\times$ 1 nm is used in the simulations. The Oersted field and the distribution of current density are numerically calculated by COMSOL\cite{comsol}. A single skyrmion is first created at the position $x$ = 30 nm from the left edge of the strip, then moves along strip under current-driving. Figure 1(b) shows an example of the current-driven motion of skyrmion under an ac continual square current pulse with an amplitude of $\pm$16~$\mu$A and pulse width of 14 ns. Fig. 1(c) shows that skyrmion exhibits a non-uniform motion with a strong movement velocity dependence on its current position $x$ under the driving currents.  The spin torque driving non-uniform motion of skyrmion is caused by the nonlinear change repulsive force on skyrmion from dipole field, non-uniform demagnetizing field, and current-induced spin torques, and further enhanced by non-uniform distribution of current density on dumbbell shape strip of Pt/Co. The current-driven non-uniform motion of skyrmion causes a nonlinear dependence of the position $x$ on amplitude and pulse width of the current $I$. The current pulse $I$ and $x$ are defined as the input and output of physical reservoir discussed below, respectively. The variations and nonlinear properties are beneficial to build a well physical reservoir that will be proved by the handwritten digits recognition task below.

\begin{figure}[htbp]
\includegraphics[width=0.45\textwidth]{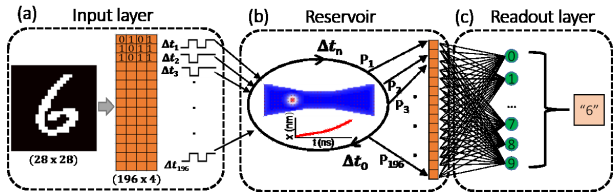}
\caption{Process flow diagram of handwritten digit recognition using an MSM-based RC system. (a) Signal preprocessing in the input layer: the digit image was first converted to a 196 by 4 matrix, and each row of the matrix was encoded by a 4-bits current pulse, and in time order was fed into the MSM of RC system. (b) Reservoir: A vector consisted of 196 skyrmion positions $x$ was obtained after signal processing by an MSM-based reservoir because only the last position $x$ was recorded after each 4-bits current pulse. (c) A linear readout layer: The recognition result was acquired after linearly mapping the reservoir states to a trained readout function.}\label{fig2}
\end{figure}

\subsection{B. Spin torque nano-oscillator}

Fig. 1(d) shows the block diagram of signal detection and schematic of the modeling STNO consisted of two magnetic layers (fixed and free) and one barrier layer. The material parameters used in the simulations are the exchange stiffness $A$ = 13 pJ/m, the saturation magnetization $M_{s}$ = 800 kA/m, the Gilbert damping constant $\alpha$ = 0.01, and the spin polarization $P$ = 0.4. The cell size of 5 nm $\times$ 5 nm $\times$ 5 nm is used in the simulations. The Oersted field is also considered in the simulation. The stable magnetization oscillation is obtained at current $I$ varying from 0.01 to 0.40 mA with a free external field. Figure 1(e) shows an example of the time dependence of the normalized x-component of magnetization $m_x/m$ of the free layer on a square current pulse of 0.2 mA and 50 ns. The magnetization oscillation shows a current dependent relaxation matching process with a relaxation time $\tau(I)$ after sharply switching current pulse. Fig. 1(f) shows that the maximum envelope of $m_x/m$ procession motion has a nonlinear dependence on the amplitude of the current $I$. The current pulse $I$ and $m_x/m$ are defined as the input and output of physical reservoir discussed below, respectively.

\section{III. Results of physical reservoir systems}

\subsection{A. Handwritten digits recognition}

\begin{figure}[htbp]
\centering
\includegraphics[width=0.45\textwidth]{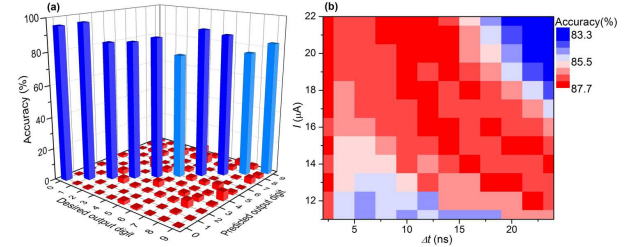}
\caption{Results of handwritten digit recognition using a RC system with one signal MSM. (a) Handwritten digit recognition rates is presented in the plane of the desired outputs and the predicted values with using current pulses of 20 $\mu$A and 10 ns. (b) Pseudocolor mapping of the recognition accuracy rates in the plane of current pulse amplitude $I$ and width $\Delta t$.}\label{fig3}
\end{figure}

 A reservoir computing system consists of three essential layers: an input layer, the reservoir, and a readout layer\cite{JaegerScience2004,prl7}. In the input layer, the original signals need to do such identity or bias transformations and be transformed to the final input vectors, also called as the preprocessing of signals. As the reservoir layer, the physical reservoir is no need to be trained due to the fixed nature of the weights between neurons in the reservoir itself. The readout layer must be sufficiently powerful to map the output vectors from the reservoir to desired output results with a good wight obtained via training. Here, we use handwritten digits recognition as an example to illustrate a physical RC system and its computing procedure at Fig.2. The Mixed National Institute of Standards and Technology (MNIST) database is a standard machine learning dataset\cite{mnist}. The original handwritten digit image in MNIST with 28$\times$28 pixels is the greyscale image. To simplify the input data, the original greyscale image was first converted into a black and white two-value image with the same pixels, then transformed into a 196 $\times$ 4 matrix, and each row of matrix as a unit was encoded by a 4-bits current pulse with specific value offset (or bias), as an input vector. After the transformation, the two-dimensional image are converted in row-wise (or column-wise) operation into temporal input to the reservoir [Fig.2(a)]. It should be noted that only last position $x$ needs to be recorded after each 4-bits current pulse. Therefore, the temporal input with 196 vectors (each vector including four current pulses) is separately fed into the MSM reservoir, and which generates an output vector $p(x)$ including 196 positions of skyrmion [Fig.2(b)]. Finial, the recognition result was produced by linearly mapping the output vector presenting the reservoir state to a trained readout layer[Fig.2(c)].

The handwritten digits dataset is extensively studied in testing new neural networks. The handwritten digits recognition is chosen as a test task to evaluate the performance of our RC system built using one single MSM. As mentioned before, a handwritten digit image with 28$\times$28 pixels is first converted to a temporal input including 196 time-series sequence streams, and each stream consists of 4-bits binary current pulse where the binary digits 0 (-20 $\mu$A, 10 ns) and 1 (20 $\mu$A, 10 ns) correspond to the black and white pixel of image respectively. In principle, the position state of the MSM will be changed after applying a current pulse, and a 4-bits current pulse stream can represent 2$^4$ different current states and then generates corresponding 2$^4$ memristor states (positions $x$) after stimulation. Similar to the prior memristors-based RC system for pattern recognition\cite{DuNC2017}, we take the same approach that the temporal input is separately fed to an MSM in 4 current pulses as a stream and only last position $x$ of skyrmion is recorded after each stream stimulation [See Fig.S2 in SM]\cite{supp}. Besides, MSM is refreshed to the initial state before feeding the next stream with 4-bits pulse. The MSM refresh operation aims to make us be able to deal with ten thousands of images based on 2$^4$ independent the simulation results for each amplitude or width of a pulse. It should be pointed out that the refresh operation is not necessary for the practical application of MSM-based RS system in pattern recognition. Therefore, the temporal input taking a specific feature of handwritten digit image is represented as a position $x$ feature vector with 196 elements in the reservoir, which can be used to perform pattern recognition through the readout layer after training.

In the readout layer, the output feature vector with 196 elements is linear all connection with the ten output results (digits 0-9). The readout layer does all the learning and training using gradient descent method\cite{LeCunIEEE1998} with the $sigmoid$ function to find the optimal weight matrix of 196$\times$10 and ten bias values. The readout function was achieved by using Python software\cite{supp}. Fifty thousand images from the MNIST data set was used for the output layer training. An optimal weight matrix and ten bias values are found and fixed after training, and another set of samples consisting of 10000 images not used in the training set, are used to test image recognition accuracy.

The recognition results obtained from the RC system with an MSM $vs.$ the desired outputs are shown at Fig.3(a). It should be noted that nonlinear dynamics and memory behaviors of magnetic skyrmion memristor strongly depend on applied current, discussed early in Fig.1(c), which should have significate influence on pattern recognition. To get the optimal operating conditions for pattern recognition, classification of handwritten digit images is performed systematically in a wide range of the amplitude $I$ and pulse width $\Delta t$ of current. As can be seen in Fig.3(b), the better accuracy of pattern recognition is located at the diagonal area of the I-$\Delta t$ plane, indicating when skyrmion locates at near the center of dumbbell stripe, the MSM has a better current-induced dynamic variation and fading memory meaning the coming position $x$ of skyrmion depends on the past and present $x$. The best recognition accuracy of 87.6$\%$ is achieved at applied current with amplitude $I$ = 20 $\mu$A and width $\Delta t$ = 10 ns, comparable to other results $88\%$ on simulation of an one-layer neural network with 7850 parameters\cite{LeCunIEEE1998}. The recognition rate can be much improved if the diversity (variable states) of the reservoir is further increased by adding more different MSMs for reservoir or using more than one different pulse width, commonly used in prior RC systems\cite{DuNC2017}.

\subsection{B. Solving unknown complex nonlinear dynamic systems}

\begin{figure}[htbp]
\centering
\includegraphics[width=0.45\textwidth]{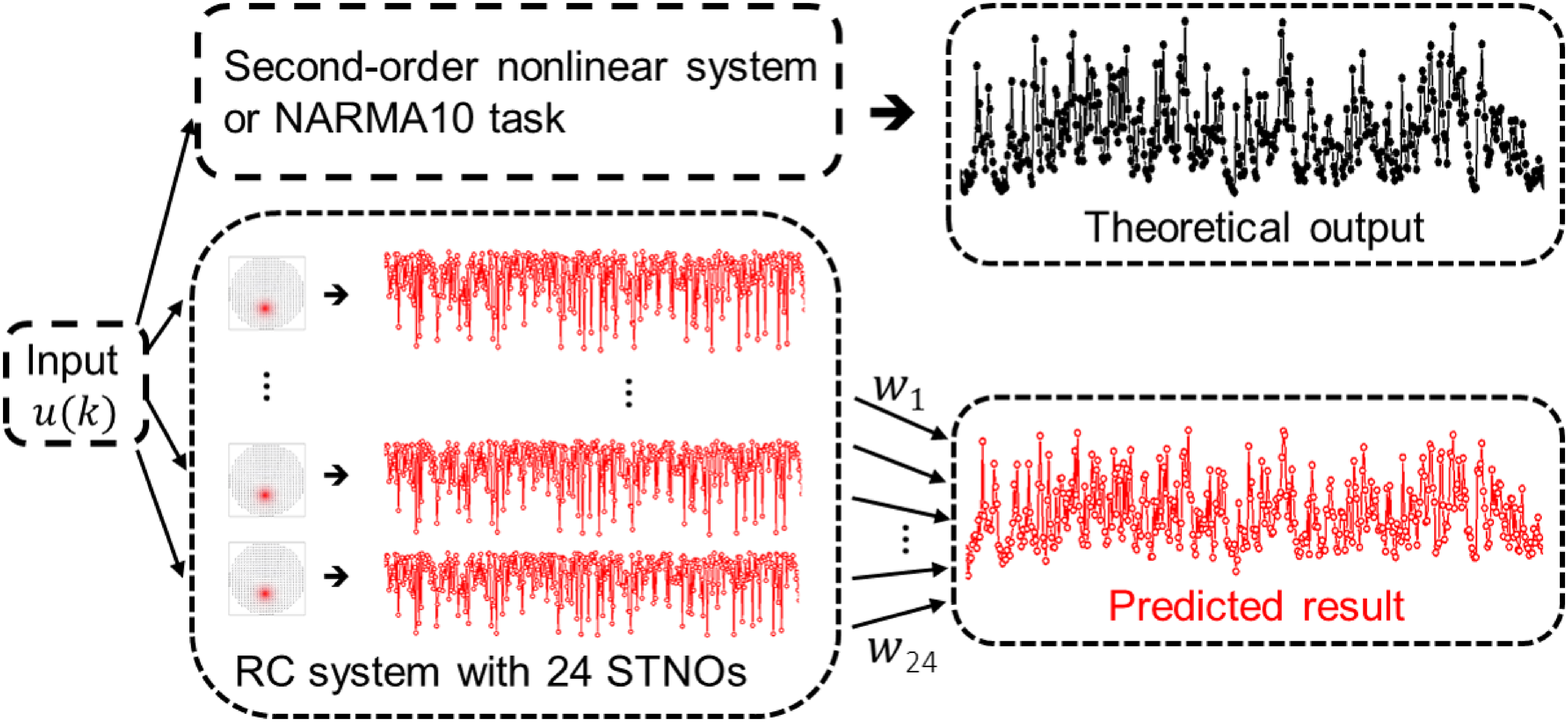}
\caption{Process flow diagram of solving unknown nonlinear dynamic systems using a RC system with 24 STNOs. The theoretical output signal $y(k)$ is obtained by putting uniform random input signal $u(k)$ to the original nonlinear dynamic system described by Eq.(1) (or Eq.(2)) in text (upper panel). For neuromorphic processing, the same random input signal was encoded by 24 different current pulse streams, and then were separately fed to 24 STNOs of the RC system. The 24 output curves generated from 24 STNOs of RC system were summed as a predicted output $y_{p}(k)$ by linear superposition with 24 factors ($w_i$ = 1,...24) (down panel).}\label{fig4}
\end{figure}

In pattern classification, an spatial pattern is first partitioned by row-wise (or column-wise) into temporal (time sequence) signals to the reservoir. In fact, the reservoir computing system is more natural and suitable to deal with time series system\cite{prl6,prl7,JaegerScience2004}, such as reproducing nonlinear dynamics of an unknown complex system and speech recognition\cite{DuNC2017,AppeltantNC2011,TorrejonNature2017}. Based on the nonlinear dynamics of MSM and STNO, discussed early in Fig.1 and Fig.3S in Supplemental Materials (SM)\cite{supp}, STNO has more current-dependent dynamics diversity than MSM. In addition, STNO device has been intensively studied in experiment and theory for decades and has a standard preparation process to be achieved in both academic and industry\cite{PribiagNaturePhysics2007}. Below, we use both of a second-order nonlinear dynamic system and a Nonlinear Auto-Regressive Moving Average equation of order 10 driven by white noise (NARMA10) to test our physical RC system consisted of 24 STNOs.

\subsubsection{B1. Solving a second-order nonlinear dynamic task}

\begin{figure}[htbp]
\centering
\includegraphics[width=0.45\textwidth]{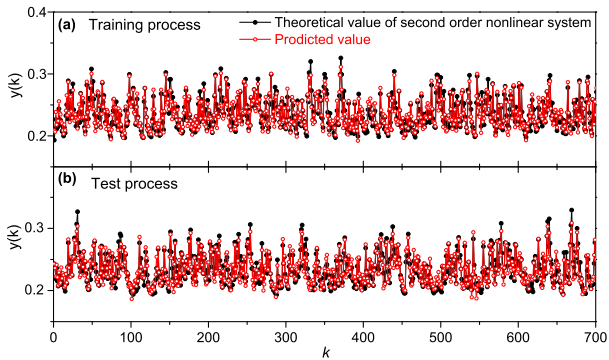}
\caption{The second order nonlinear dynamic system described in Eq.(1). (a) Theoretical output (black solid circle) vs the reconstructed output from the RC system (red circle) in training phase. (b) Same as (a) in test phase.}\label{fig5}
\end{figure}

Nonlinear systems are ubiquitous in the real world. Second-order nonlinear dynamic systems are widely used to model systems across the science and engineering, especially in the electrical system because they are the simplest type of dynamic system to exhibit oscillations such as mass-spring-damper systems and RLC circuits. In addition, many true higher-order systems may be approximated as second-order to facilitate analysis. To directly compare with the RC system bases on metal-oxide memristors\cite{DuNC2017}, the same second-order nonlinear dynamic system in the prior study is chose to test our physical RC system. The nonlinear dynamic system generates an output $y(k)$ following a second-order nonlinear transfer function defined as follows:

\begin{equation}
\begin{split}
y(k) = 0.4*y(k-1) + 0.4*y(k-1)y(k-2) \\
+ 0.6*u(k)^3 + 0.1
\end{split}
\end{equation}

The Eq.(1) shows the output $y(k)$ depends on both of the past two outputs $y(k-1), y(k-2)$ and the current input $u(k)$ and the relationship between $y(k)$ and $u(k)$ is implicit and hidden. To deal with this kind of nonlinear problems, it needs an RC system with enough varieties as well as a fading memory. As shown in Fig. 1, STNO exhibiting complex nonlinear dynamics and good short-term memory (relaxation time $\tau \sim$ 25 ns) is expected to have outstanding performance in solving nonlinear problems\cite{supp}. Figure 4 illustrates the process flow diagram of the STNOs-based RC system dealing with dynamic time-series data. The reservoir consists of 24 STNOs including three different diameters (240, 270 and 300 nm). The same uniform random time-series input sequence with a length of 800 points used in the theoretical Eq.(1) (or Eq.(2)) was firstly scaled to 24 current pulses sequences to stimulate 24 STNOs of RC system [See Table.I in SM]\cite{supp}. To increase the diversity of reservoir, 24 input current pulse sequences with 13 different amplitudes or pulse widths are used to stimulate 24 STNOs separately and generated 24 independent output curves including 800 time-series dynamic states in each curve from 24 STNOs of RC system [See Fig.S6 in SM]\cite{supp}. As the same as the handwritten digits recognition, we still chose to implement a linear regression using gradient descent to find 24 weighed factors ($w_i$ = 1,...24) and one bias value during the training process (See detail in SM\cite{supp}).

Fig.5(a) shows the simulation output reconstructed from the RC system after training and the theoretical output from the training sequence, indicating the RC system with 24 STNOs can correctly solve the nonlinear dynamic problem, with a normalized mean squared error (NMSE) of $1.17*10^{-3}$ in training phase. To verify the physical STNOs-based RC system has indeed solved the second-nonlinear dynamic transfer function, a new random sequence independent with the training sequence as input to test the RC system. The result of Fig.5(b) demonstrates that the system still successfully predict the expected dynamic output for the new random sequence using the same weight matrix, with a similar NMSE of $1.31* 10^{-3}$. Now it is safe to say that the RC system has solved the second-nonlinear dynamic transfer function after training an 800-long history time-series data. Moreover, the RC system with 24 STNOs has better performance than the RC system with 90 metal-oxide memristors (NMSE $\simeq$ $3.13*10^{-3}$) in prior reports\cite{DuNC2017}

\subsubsection{B2. Solving NARMA10 nonlinear dynamic task}

To further explore the capabilities of the STNO-based RC system in solving dynamic nonlinear problems and analyzing time series data, NARMA10 as another nonlinear dynamic system was chose to test our RC system. The NARMA10 model is widely used to simulate time series and as a benchmark in the reservoir computing community\cite{AppeltantNC2011}. The NARMA10 model is given by the recurrence formula as following:

\begin{equation}
\begin{split}
y(k) = 0.3*y(k-1) + 0.05*y(k-1) \sum_{i=1}^{10}y(k-i) \\
 + 1.5*u(k-1)u(k-10) + 0.1
\end{split}
\end{equation}

where $u(k)$ is a sequence of random inputs with a uniform distribution between 0 and 0.5. Comparing to the second-order nonlinear dynamic problem described by Eq.(1), the NARMA10 has more complex and distinctive nonlinear dynamic features because its output $y(k)$ not only depends on the past ten outputs $y(k-1)$, ... and $y(k-10)$ but also is related to the cross term of the past inputs $u(k-1)$ and $u(k-10)$. The same RC system with 24 STNOs used in the second-order task still can well capture the most of nonlinear dynamic feature of NARMA10 after training an 800-long history time-series data. Fig.6 shows that the outputs reconstructed from the RC system after training process well match the theoretical outputs. The performance is evaluated by the normalized root mean square error (NRMSE) between predicted output and theoretical output. NRMSE is 0.128 (or 0.123) in the test phase (training phase), which is better than NRMSE$\simeq$0.15 in a digital reservoir of 400 nodes (400 parameters needed to train)\cite{AppeltantNC2011}.

\begin{figure}[htbp]
\centering
\includegraphics[width=0.45\textwidth]{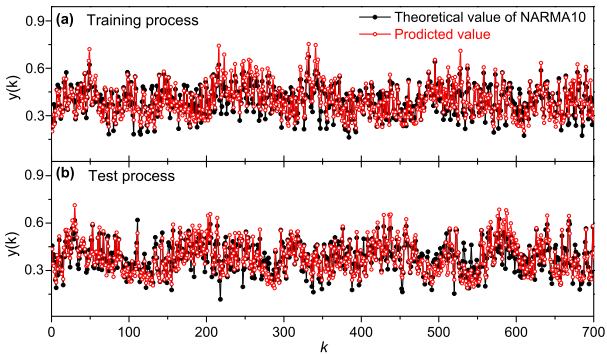}
\caption{NARMA10 task described in Eq.(2). (a) Theoretical output (black solid circle) vs the reconstructed output from the RC system (red circle) in training phase. (b) Same as (a) in test phase.}\label{fig6}
\end{figure}

\section{IV. Conclusions}

Based on the complex nonlinear dynamic and recurrence behaviors of spintronic nanodevices, we proposed two physical reservoir computing systems based on magnetic skyrmion memristor and spin-torque nano-oscillator, respectively. Since the nonlinear magnetic dynamics of spin-torque nanodevices can be very well described and captured by numerically solving the LLG equation using micromagnetic simulation software, ie., OOMMF, the physical RC system based ST devices can be analyzed through micromagnetic simulation. The performance on temporal information processing of both of physical RC system using an MSM and using 24 STNOs (indeed three different STNOs) are evaluated through performing handwritten digits recognition and solving second-order nonlinear dynamic problems and NARMA10, and the three tasks have been widely used at the benchmark in the reservoir computing community.

Our results demonstrated that the RC system with only one MSM could achieve a comparable accuracy $87.6\%$ of pattern recognition as the reservoir with 88 metal-oxides memristors, meanwhile the information processing using MSM-based RC system has more than 1000 times fast than metal-oxides memristor-based RC system due to magnetization dynamics at the nanosecond time scale compared to the motion of ions at the millisecond time scale\cite{DuNC2017,PreziosoNature2015,Sheridan2017}. Furthermore, compared to the previous reports, second-order and NARMA10 nonlinear system tasks demonstrated STNOs-based RC system had outstanding performance in solving unknown and complex nonlinear dynamic problems. It should be noted that physical RC system is robust to the actual device defects and deviation (in fact, certain diversity is necessary) due to the fixed nature of the weights between neurons in the reservoir itself. Our simulation results indicate that spin-torque nanodevices are very suitable to build various physical reservoirs or construct an artificial neural network for neuromorphic computing, and provide some ideas and strategies for experimentally building physical reservoir computing system.

\section{Acknowledgements}
L.N.C, Y.W.D and R.H.L are supported by from National Key Research and Development Program of China (2016YFA0300803), National Natural Science Foundation of China(No.11774150), Applied Basic Research Programs of  Science and Technology Commission Foundation of Jiangsu Province (No.BK20170627), and the Open Research Fund of Jiangsu Provincial Key Laboratory for Nanotechnology.

\end{document}